\documentclass[fleqn,12pt]{wlscirep_mbk}
\usepackage[utf8]{inputenc}
\usepackage[T1]{fontenc}

\usepackage{graphicx}	
\usepackage{amsmath}	
\usepackage{amssymb}	
\usepackage{aas_macros}
\usepackage{xcolor}
\usepackage{enumitem}
\usepackage[symbol]{footmisc}

\usepackage{url}

\usepackage{color}
\usepackage{hyperref}
\definecolor{linkcolor}{rgb}{0,0,0.5}
\hypersetup{colorlinks=true,linkcolor=linkcolor,citecolor=linkcolor,filecolor=linkcolor,urlcolor=linkcolor}

\newcommand{\msun}{${\rm M}_{\odot}$}
\newcommand{\zsun}{${\rm Z}_{\odot}$}
\newcommand{\mstar}{${\rm M}_{\star}$}

\newcommand{\hii}{{\sc HII}}

\title{Near-Field Cosmology with the Lowest-Mass Galaxies}

\author[1]{Daniel R.\ Weisz}
\author[2]{Michael Boylan-Kolchin}
\affil[1]{Department of Astronomy, University of California Berkeley, Berkeley, CA 94720, USA; dan.weisz@berkeley.edu}
\affil[2]{Department of Astronomy, The University of Texas at Austin, Austin, TX 78712, USA; mbk@astro.as.utexas.edu}

\whitepapertext{Submitted in Response to the Astro2020 Call for Science White Papers}

\begin{abstract}
The term ``near-field cosmology'' broadly describes connections between the low- and high-redshift Universe.  Resolved studies of $z=0$ galaxies less massive than the LMC (\mstar~$\lesssim 10^{9}$ \msun) provide access to (i) small scales of the matter power spectrum that are challenging to study otherwise and (ii) the galaxy population that is fainter than high-redshift direct detection limits.  \textbf{The study of nearby low-mass galaxies has implications that extend far beyond the local Universe and include the nature of dark matter, cosmic reionization, and galaxy formation across cosmic time.}
\vspace{0.15cm}

\textbf{Dark Matter:}~~The only evidence for dark matter's existence comes from its inferred gravitational influence on galactic and cosmological scales. Near-field cosmology provides our best evidence for the amplitude of density fluctuations on sub-galactic scales: so long as ultra-faint galaxies are hosted by dark matter halos, their existence and inferred abundance require the matter power spectrum to extend down to mass scales of $M \sim 10^{9}$~\msun. Kinematic information from large samples of individual line-of-sight and transverse velocities of stars and from spatially resolved gas kinematics in nearby dwarf galaxies directly informs our understanding of dark matter interactions in the deeply nonlinear regime. 

\vspace{0.15cm}
\textbf{Galaxy and Stellar Astrophysics}:~~Resolved stars in low-mass galaxies serve as crucial and unique foundations for interpreting the light of galaxies at all cosmic epochs, including faint galaxies in the distant Universe.  
Fundamental measurements include star formation histories (SFHs) from color-magnitude diagrams, chemical abundance patterns from stellar spectroscopy, and stellar and galaxy velocities from spectroscopy and proper motions. Observations of resolved low-metallicity massive stars and HII regions are central to interpreting faint galaxies at all earlier epochs.

\vspace{0.15cm}
\textbf{Outlook:}~~In the 2020s, hundreds of new low-mass galaxies will be discovered and characterized with LSST, HST, JWST, WFIRST, and the ELTs. In the longer term, fully realizing the promise of near-field cosmology with low-mass galaxies requires: (i) a large aperture ($\gtrsim 9$m) UV/optical space telescope. This is the only way to measure SFHs and proper motions, as well as acquire invaluable UV spectroscopy of stars and star-forming regions, to several Mpc; (ii) a two-fold community investment in ground-based optical spectroscopy: ELTs, which enable high-fidelity spectroscopy of faint and/or crowded stars; and a dedicated wide-field spectroscopic survey on a 10m class telescope.  Combined, they will uncover primordial enrichment events and signatures of the first stars, provide unique insight into dark matter, and anchor our interpretation of the faint galaxy frontier at high redshifts; (iii) radio facilities that can measure high-precision, resolved rotation curves in order to shed light into the nature of dark matter.

\vspace{0.15cm} 
\textbf{HST:}~~Finally, we suggest that \textbf{a servicing mission to extend the life of HST beyond 2025 should be seriously explored}. 
This may be a cost-effective way to bridge the looming gap in space-based UV/optical capabilities during development of a next-generation space telescope.  
Its high over-subscription rate suggests that HST can continue to enable ground-breaking science across all of astrophysics for quite some time. 
\end{abstract}

\begin{document}

\flushbottom
\maketitle

\section{Dark Matter}

\subsection{Matter Power Spectrum}
The matter power spectrum -- essentially, the contribution to the variance in the smoothed initial dark matter density field as a function of comoving smoothing scale
-- contains information about the nature of dark matter: any property of dark matter that reduces power on a particular scale will be reflected in the abundance and structure of dark matter halos at a corresponding mass \cite{bullock2017, buckley2018}. On scales comparable to the LMC and above (dark matter masses of $\gtrsim 10^{11}$~\msun~or comoving sizes of $\gtrsim 1\,{\rm Mpc}$), the power spectrum is strongly constrained to be virtually indistinguishable from the baseline $\Lambda$CDM model. 
Tests on smaller scales are much more difficult. \textbf{The near field provides our best evidence for the amplitude of density fluctuations on sub-galactic scales:} the number of known ultra-faint dwarf galaxies (UFDs) requires the power spectrum to extend to mass scales of $\sim 10^{9}$~\msun\cite{bullock2017, jethwa2018}. The total count of \mstar~$\lesssim~10^5$~\msun~dwarf galaxies within the virial volume of the Milky Way (MW) is highly uncertain (at the factor of 10 level \cite{tollerud2008,kim2018,kelley2018}) owing to spatial clustering of satellites, incomplete sky coverage, and luminosity bias. Even with a complete, stellar-mass-limited census of MW satellite galaxies, the matter power spectrum may be impossible to accurately measure because of the uncertainties related to the disruptive effects of the Galaxy's disk\cite{brooks2014,garrison-kimmel2017}. 
Observations of regions where such effects are irrelevant is therefore a high priority. \textbf{An accurate and complete census of very faint galaxies $\sim 500-1000$~kpc from the MW, which is achievable with LSST, has the potential to constrain the dark matter power spectrum on mass scales of $M \sim 10^7-10^8$~\msun.} 

\begin{figure}[b!]
\vspace{-0.4cm}
\includegraphics[width=\columnwidth]{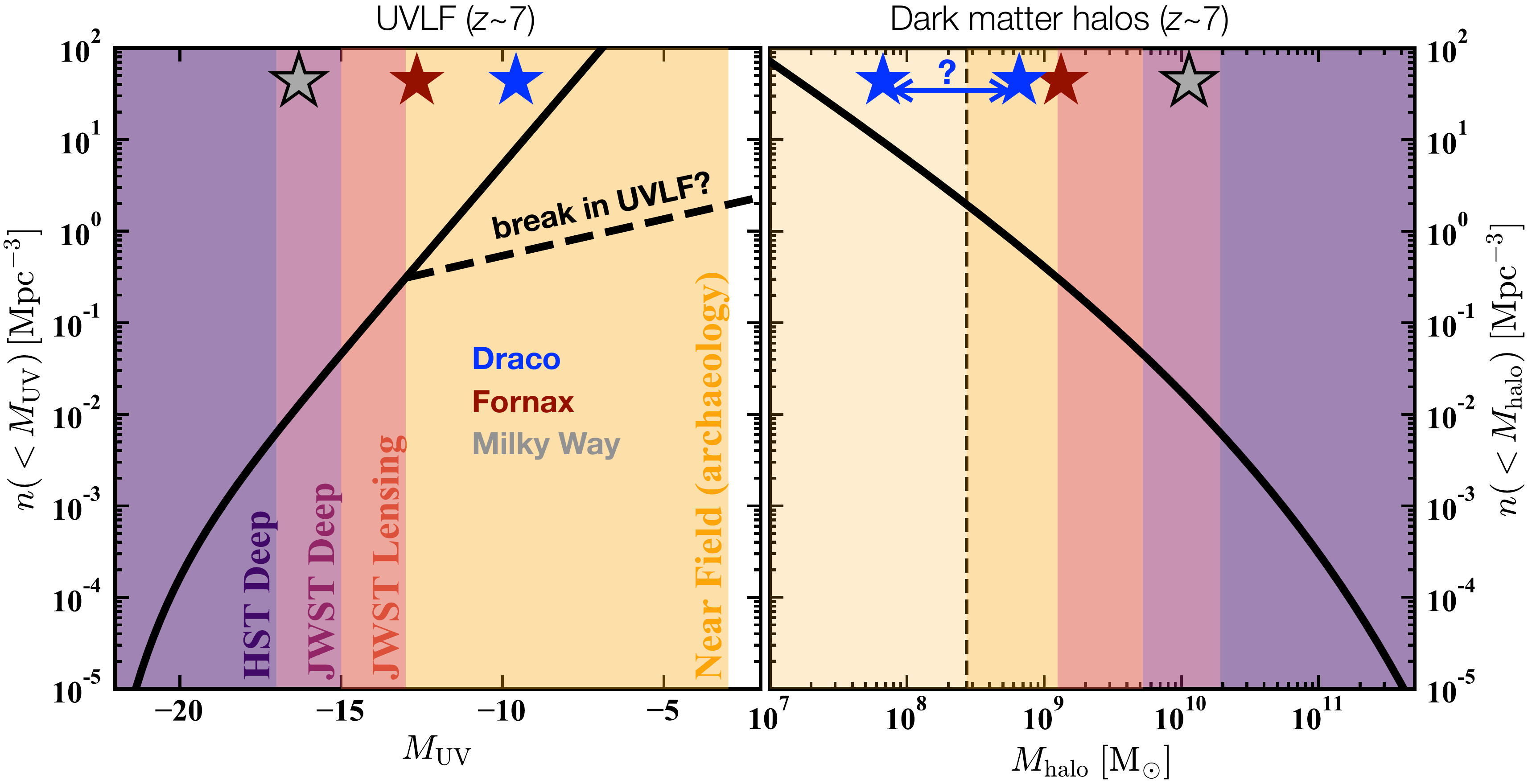}
\label{fig:halo-galaxy-connection}
{\small\baselineskip0.04cm{\hrule\vspace{0.05cm}\noindent{\sc{Figure 1:}}
The cumulative abundance $n$ of rest-frame-UV-selected objects (\textit{left}) and dark matter halos (\textit{right}) at $z\sim7$. Matching objects at fixed $n$ provides an association between galaxies and their host dark matter halos, linking galaxy observations with structure formation and dark matter physics. In this view, current blank field HST surveys are sensitive to dark matter halos with $M(z\sim7)\gtrsim 10^{10}$~\msun, while JWST+lensing will probe masses an order of magnitude lower. Archaeological reconstructions based on near-field data provide a window to  substantially lower-mass dark halos at high redshifts that may \textit{never} be accessible through direct observation. In particular, the possibility of a turn-over or break in the UV luminosity function (UVLF) will be very difficult to detect robustly even with JWST lensing fields. A census of galaxies like Draco in the near field would differentiate between the two, as an unbroken UVLF would result in an order of magnitude more local descendants than a broken UVLF. \smallskip \hrule}}
\end{figure}

Any test of the dark matter power spectrum on even smaller mass scales requires the detection of perturbations or virialized halos that will never host a galaxy. This is highly challenging, as CDM halos have densities that are extremely low relative to the typical densities of astrophysical objects: a virialized halo at $z=0$ with a mass of $M \sim 10^{5}$~\msun\ has a virial radius of $\sim1\,{\rm kpc}$, while a typical globular cluster has a similar mass but a size of $\lesssim 10\,{\rm pc}$. Nevertheless, multiple lines of inquiry in the near field may be sensitive to the power spectrum at scales below the threshold of galaxy formation. Precision measurements of density variations in globular cluster streams may reveal the presence of low-mass substructure\cite{carlberg2009a,yoon2011}, 
while the detection of wide stellar binaries at very large separations may rule out its existence\cite{penarrubia2016}. 
These measurements must be complemented with other techniques far outside of the MW (e.g., gravitational lensing\cite{dalal2002,vegetti2010}), where environmental and baryonic effects will be diminished and systematic uncertainties will be different. \textbf{Given the complete absence of non-gravitational detections of dark matter, all possible astrophysical lines of inquiry that can confirm or exclude the existence of dark matter structure below the scale of (the smallest) galaxies are essential and should be pursued vigorously.}

\subsection{Properties of Individual Halos}
 While the matter power spectrum is sensitive to dark matter physics operating in the linear regime of density perturbations, the structure of collapsed dark matter halos can be sensitive to dark matter interactions that occur when densities are high. The low baryonic content of dwarf galaxies makes them an important testing ground for CDM and its alternatives; however, results from the past several years have made it clear that comparisons between observations and CDM predictions must take into account 
 the impact of baryonic physics on the distribution of dark matter within dwarf galaxies\cite{brooks2014, dicintio2014,onorbe2015, read2016}. These effects can be grouped into two classes: ``environmental", involving any dwarf galaxies that interact with a larger system (such as the inner MW satellites interacting tidally with the MW disk and moving through its gaseous halo) and ``internal", involving energy and momentum input from star formation and stellar evolution\cite{dekel1986} (with other sources such as black holes also being potentially important).

Theoretical and computational models indicate that certain systems are preferred candidates for testing the nature of dark matter. UFDs are expected to have fewer stars per unit dark matter halo mass, meaning stellar feedback should be relatively less important\cite{tollet2016, fitts2017}. Low-mass galaxies that formed all of their stars early (before $z\sim 2$) retain their primordial dark matter density profiles in CDM+baryon simulations\cite{read2019}. Those that have never come within $\sim 50$~kpc of the MW are likely to have avoided strong environmental effects\cite{garrison-kimmel2017}. \textbf{These conditions point to faint (\mstar~$\lesssim 10^6$~\msun), distant ($\gtrsim 500$~kpc), isolated dwarf galaxies in the Local Group (LG) as ideal probes of dark matter density profiles (and therefore, of the nature of dark matter)}. While only a handful of such galaxies are known, all current evidence points to the existence of a large population lurking just below current detection thresholds\cite{simon2019}. It will be crucial to have multiple kinematic tracers in these systems -- HI rotation curves and line-of-sight (LOS) velocities of individual stars -- to mitigate uncertainties.

\section{Extra-Galactic Archaeology with Resolved Stars}

\subsection{Star Formation Histories}

 The colors and absolute magnitudes of resolved stars, along with their relative densities on various parts of a color-magnitude diagram (CMD), encode a galaxy's star formation history (SFH; which includes its age-metallicity relationship).  Techniques to recover SFHs from CMDs require minimal assumptions about the functional form of the SFH and been extensively vetted over the past few decades\cite{tosi1989, tolstoy1996, dolphin2002, skillman2002, brown2004b, hidalgo2009, monelli2010b}.

CMDs that reach the oldest main sequence turnoff (MSTO) with a SNR$\gtrsim 5-10$ per star provide the most secure SFHs\cite{cole2014, gallart2015}.  However, even HST cannot produce CMDs that reach the oMSTO beyond the LG due to its faintness (M$_{\rm V}\sim{+4}$) and crowding; SFHs from shallower CMDs are less certain\cite{weisz2011a, dolphin2012} (Figure 2).
 
Oldest MSTO-based SFHs exist for a few dozen dwarf galaxies in the LG.  Among MW satellites, the lowest-mass systems generally stop forming stars in the very early Universe\cite{brown2014}, while more luminous systems continue forming stars until later times\cite{weisz2014a}.  However, it remains unclear if this trend also exists among M31 satellites\cite{skillman2017}. 
Isolated dwarfs exhibit diverse SFHs at fixed present-day properties.\cite{skillman2014, gallart2015}.  
\textbf{Dozens to hundreds of oMSTO-based SFHs of isolated dwarfs are needed to fully determine the variance and identify trends. Achieving such sample sizes requires a $\gtrsim$9m optical space telescope (Figure 3).}

Proper motion measurements are essential to our understanding of SFHs in satellite galaxies. For example, the HST-based proper motion measurement for Leo~I\cite{sohn2013} demonstrated that Leo~I quenched at a time coincident with its pericentric passage of the MW, consistent with efficient ram pressure stripping \cite{wetzel2015c, fillingham2016}.  
 
Between HST, Gaia, WFIRST, and JWST, proper motions measurements, and similarly detailed insight into quenching, are possible for all low-mass galaxies in the LG\cite{kallivayalil2015, fritz2018, simon2018}. \textbf{Proper motions outside the LG require facilities with better angular resolution than JWST and decade-plus time baselines.}

\begin{figure}[t!]
\vspace{-0.6cm}
\begin{minipage}[l]{0.5\linewidth}
\includegraphics[scale=0.5]{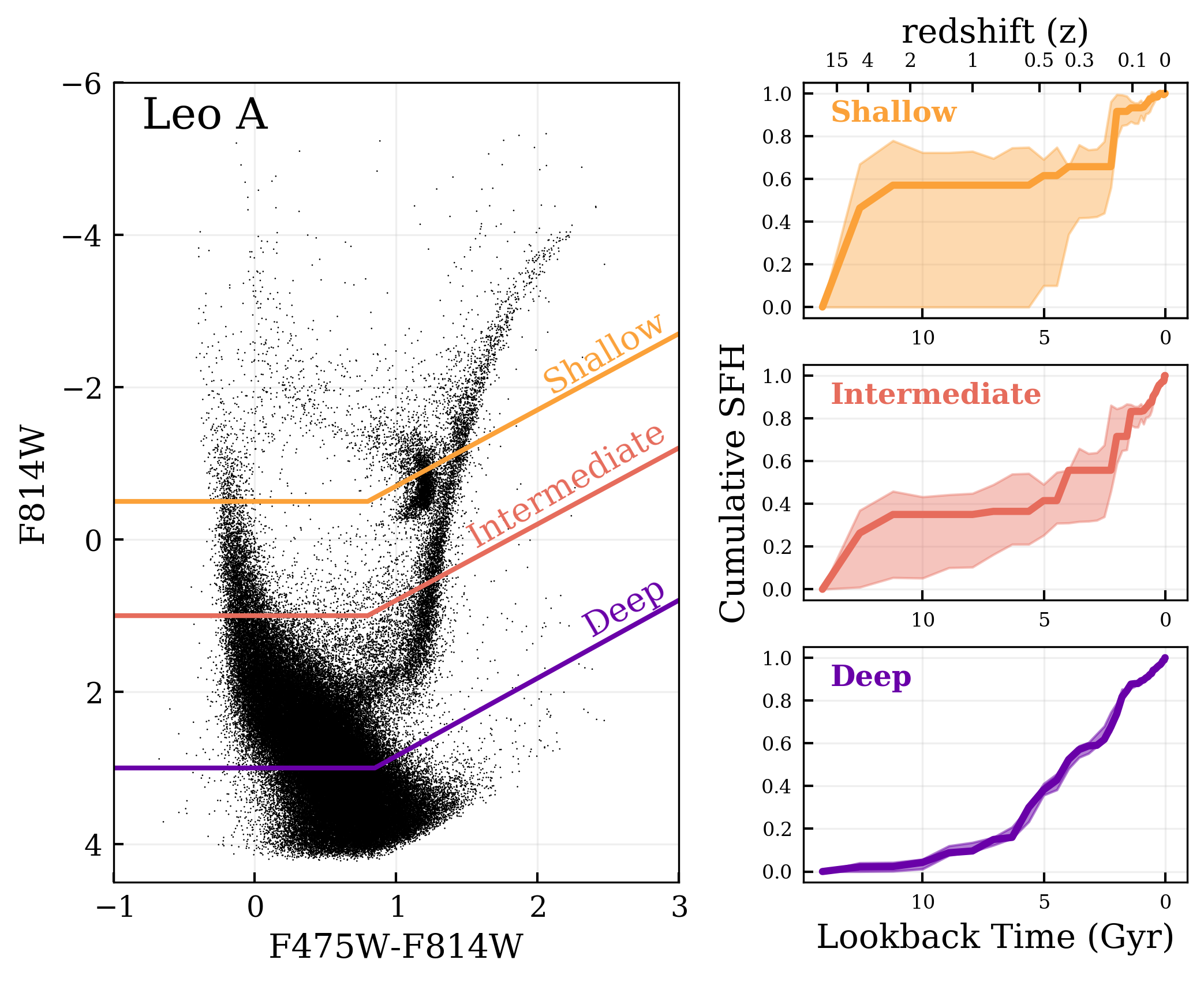}
\label{fig:}
\end{minipage}\hfill
\begin{minipage}[t]{0.48\linewidth}
\vspace{-95pt}
{\small\baselineskip0.04cm{\hrule\vspace{0.05cm}\noindent{\sc{Figure 2:}}
{\textbf{Left:} The HST-based CMD of LG dwarf irregular Leo~A\cite{cole2007} ($D\sim800$~kpc, $\log(M_{\star}/M_{\odot}) \sim 6.8$, $\log(Z_{\star}/Z_{\odot}) \sim 10\%$)\cite{mcconnachie2012}.  The colored lines represent the variety of CMD depths that are typical of archival HST data in and beyond the LG.  HST can only reach the oldest MSTO within the LG due to sensitivity and crowding limitations. \textbf{Right:} The cumulative SFHs (fraction of stellar mass formed prior to a given epoch) of Leo~A measured from the CMD (at different depths) in the left panel. Uncertainties include contributions from random (e.g., number of stars\cite{dolphin2013}) and systematic components (e.g., from variations in stellar models\cite{weisz2011a, dolphin2012}).  CMDs that include the oldest MSTO provide a precise SFH measurement at all epochs, while SFHs measured from shallower CMDs are uncertain at older ages.  
\smallskip \hrule}}}
\end{minipage}
\vspace{-0.4cm}
\end{figure}

\subsection{Reionization}
The lowest mass galaxies in the high-redshift Universe are beyond the direct detection limits of our most powerful telescopes, including JWST (Figure 1).  \textbf{The study of resolved dwarf galaxies may be the only way to observationally link the faintest galaxies to reionization.} This topic is of the utmost importance for both cosmology and galaxy formation modeling\cite{Robertson:2010vq, robertson2015, mirocha2019}.

The main progenitors of classical LG dwarf galaxies such as Draco and Leo~A had UV luminosities consistent with the faint galaxy population thought to drive reionization\cite{weisz2014d,boylankolchin2015}.  Moreover, UFDs around the MW require that the high-redshift UVLF extends as faint as $M_{\rm UV}(z=7) \sim -3$\cite{weisz2017}. A break in the high-redshift galaxy UVLF appears necessary to match LG dwarf galaxy number counts\cite{boylankolchin2015, simon2019}.  This break may be impossible to observe directly at high redshifts, even with gravitational lensing (Figure 1).  \textbf{An accurate census of nearby dwarfs along with their oMSTO-based SFHs may be the only way to constrain shape of the faint UVLF in the early Universe.}

The relatively small number of low-mass satellite galaxies around the MW can be understood naturally if cosmic reionization sets a lower mass limit on galaxy formation: halos below the atomic cooling limit of $T_{\rm vir} \approx 10^4\,{\rm K}$ were unable to form stars once hydrogen was reionized\cite{bullock2000, benson2002}. The discovery of UFDs and the revelation that their stellar populations are generally consistent with no star formation since the reionization era\cite{brown2014, weisz2014a} helped to solidify the idea that the ``missing satellites problem''\cite{klypin1999,moore1999} might not be fatal for CDM.  
SFHs of some low-mass dwarf galaxies, along with cosmological simulations, now suggest that reionization may not quench all of the lowest-mass galaxies simultaneously due to inhomogeneities in reionization and/or self-shielding\cite{ocvirk2013, weisz2014m31, monelli2016, fitts2017}.  \textbf{Measuring the oMSTO-based SFHs for a large number of UFDs, particularly those that are not satellites, provides a unique test of reionization scenarios. }

\subsection{Chemical Enrichment and the First Stars}

The low metallicities of resolved stars in dwarf galaxies imply they have been subjected to relatively little enrichment, thus providing a plausible pathway for identifying individual enrichment events. 
For example, the enhancement of r-process elements such as Ba and Eu in some UFDs suggests that these systems hosted singular neutron capture events (e.g., neutron star mergers) in the early Universe\cite{ji2016, duggan2018, marshall2018}.   Measurements of just a handful of elements, such as Fe or a few $\alpha$-elements, can be used to distinguish between leaky and closed box chemical evolution or constrain the relative timing of supernovae-based enrichment\cite{kirby2009, kirby2013}. In galaxies with purely ancient populations, such as UFDs, the abundance patterns can provide important insight into the lives and deaths of the first stars\cite{frebel2015}.

The number and quality of abundance determinations scales with distance, resolution, and SNR.  Medium resolution ($R\sim6000)$ spectroscopy can provide several elements in MW satellites\cite{kirby2008a}, and [$\langle \alpha \rangle$/Fe] with $10+$ hour times at the edge of the LG \cite{kirby2017}.  High resolution ($R \gtrsim 10,000-20,000$) spectroscopy provides dozens of elements\cite{tolstoy2009}, and is particularly information rich at UV wavelengths\cite{roederer2012}.  New spectral fitting techniques can recover 10-20 elements from low resolution ($R\sim2000$) spectra\cite{ho2016, ting2017}.  \textbf{Substantially improved knowledge of abundance patterns and enrichment processes requires a combination of wide-field narrow-band photometric\cite{keller2007, starkenburg2017} and/or modest resolution spectroscopic surveys\cite{cui2012, chiba2016} with higher resolution follow-up in the optical/UV, either from ELTs\cite{simon2019} or space-based observatories.}

\subsection{Resolved Low Metallicity Calibrators for the High-Redshift Universe}

Interpreting the spectral energy distributions and emission lines of faint, metal-poor, star-forming galaxies requires knowledge of low-metallicity massive stars and \hii\ regions.  For massive stars, resolved observations of the SMC ($Z\sim 20\%$~\zsun) have long served as the metal-poor empirical anchor\cite{leitherer1999}. However, it may not be a sufficiently metal-poor proxy for the faintest systems at high redshifts \cite{erb2010, madau2014b, stark2016,senchyna2017, berg2018}.  At lower metallicities, massive stars have different rotation rates, wind strengths, surface chemistry, and intrinsic binarity, which affects properties such as lifetimes and the output of ionizing photons\cite{levesque2012, choi2017, stanway2016}. Nebular emission in \hii\ regions is also highly sensitive to metallicity and massive stars properties\cite{conroy2013, steidel2016, byler2017, byler2018}.

Expanding local calibrations of low-metallicity massive stars and \hii\ regions is challenging.  Most sub-SMC metallicity star-forming regions are located at the periphery of the LG or beyond.  Acquiring high SNR spectroscopy of massive stars at such distances, particularly in the UV, requires large integration times with our most powerful telescopes\cite{garcia2014,garcia2019, evans2019}.  Comparatively more progress has been made on low-metallicity \hii\ regions at optical wavelengths\cite{berg2012}.  Rest-frame far-UV spectra of \hii\ regions (especially $\sim$900-1200 \AA) are particularly feature-rich\cite{heckman2011,steidel2016} but are not easily observed.  \textbf{Improving local calibrations of low-metallicity massive stars and \hii\ regions, particularly at UV wavelengths, is imperative for accurately interpreting the light of faint, star-forming galaxies in the high-redshift Universe.}

\begin{figure}[t!]
\vspace{-0.4cm}
\begin{minipage}[l]{0.58\linewidth}
\includegraphics[scale=0.58]{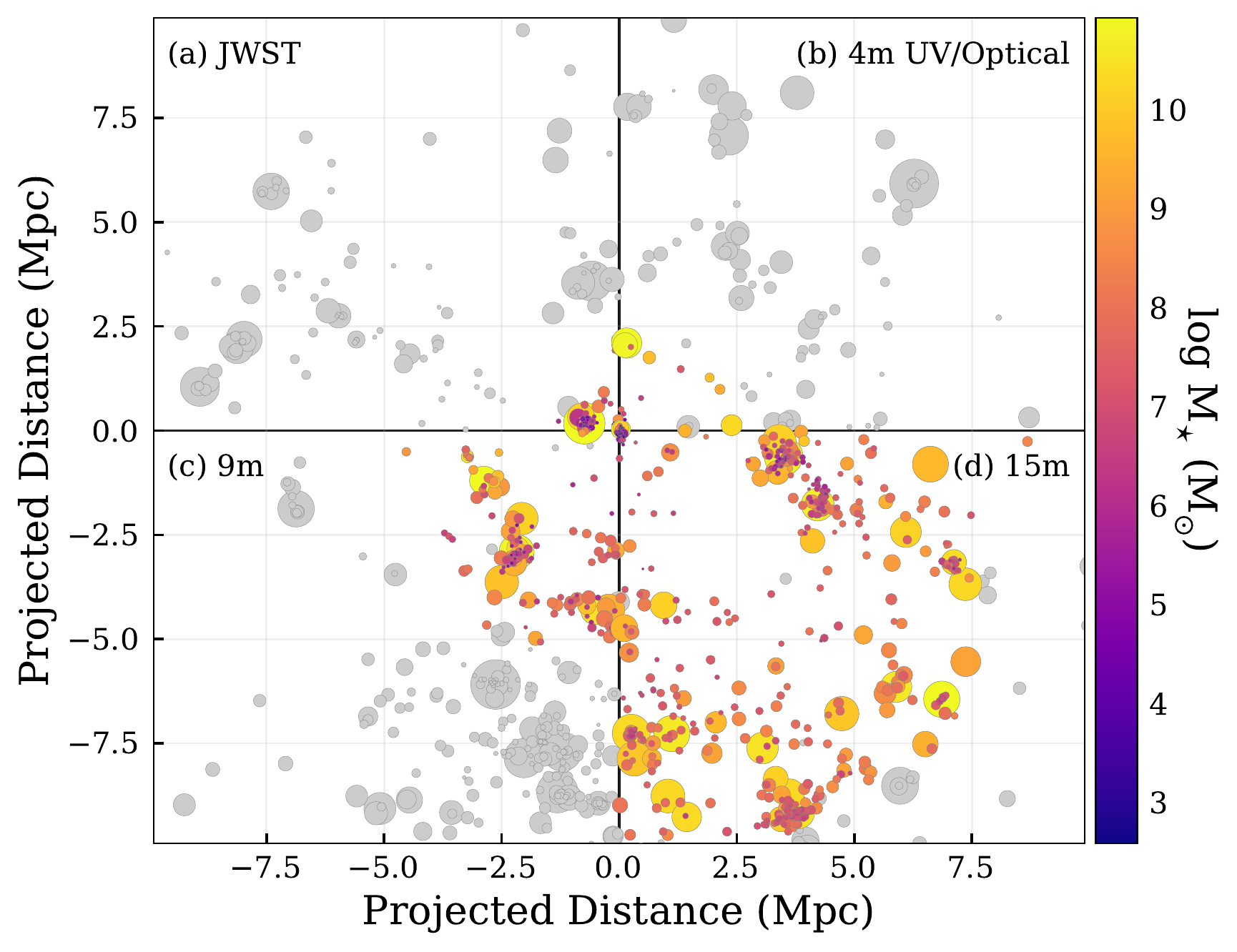}
\label{fig:lvmap}
\end{minipage}\hfill
\begin{minipage}[t]{0.41\linewidth}
\vspace{-115pt}
{\small\baselineskip0.04cm{\hrule\vspace{0.05cm}\noindent{\sc{Figure 3:}}
{A projected 2D map of known Local Volume galaxies\cite{karachentsev2013} centered on the MW. Point sizes are proportional to half-light radii.  In each quadrant, a galaxy for which a given facility can reach the oMSTO in $\lesssim$100 hours of integration time, assuming negligible crowding effects, is colored-coded by \mstar.  JWST is limited mainly by its poor blue sensitivity.  Projections for the 4, 9, and 15-m UV/optical options are adopted from the LUVOIR STDT\cite{luvoir2018}.  In total, the 4, 9, and 15m UV/optical telescopes can reach the oldest MSTO in $\sim$150 ($\sim$2.5 Mpc), $\sim$500 ($\sim$5 Mpc), and $\sim$1000 ($\sim$10 Mpc) known galaxies.  Reaching the oldest MSTO is the most demanding observation, and thus enables a range of other science (e.g., proper motions). \smallskip \hrule}}}
\end{minipage}
\vspace{-0.4cm}
\end{figure}

\section{Future Prospects}

\noindent \textbf{Spectroscopy:} Substantial progress in understanding the nature of dark matter, the first stars, and enrichment mechanisms in the early Universe requires a three-fold investment in spectroscopy.  First, medium and high-resolution spectroscopy are needed to measure LOS velocities and abundance patterns of faint stars in, for example, UFDs (e.g., discovered by LSST, WFIRST), as well as to improve our knowledge of low-metallicity massive stars and HII regions.  This requires a 30m class ground-based facility.  Second, it is important to collect large samples of stars in order to establish broad trends (e.g., kinematics vs.~abundances), identify rare populations (e.g., extremely metal-poor star),  characterize resolved ultra-diffuse dwarf galaxies\cite{torrealba2016, torrealba2018}, and improve the efficiency of follow-up with 30m telescopes (e.g., membership identification).  This can be achieved to moderately faint limits with a wide-field, highly multiplexed optical spectograph on a dedicated 10m class telescope\cite{mcconnachie2016}.   Finally, UV spectroscopy from a $\sim$9-15m class space telescope is necessary to reveal detailed abundances of ancient stars and characterize the information rich UV spectra of low-metallicity massive stars and HII regions. 
\emph{If} the HST ``UV Legacy Initiative'' proposed by STScI includes a substantial number of targets at sub-SMC metallicities,
it may help to address some of the current observational shortcomings of low-metallicity massive stars.
\vspace{0.15cm}

\noindent \textbf{Imaging:} Connecting local low-mass galaxies to cosmic reionization and high-redshift galaxy studies requires imaging the oldest MSTO for hundreds of nearby dwarf galaxies.  As shown in Figure 3, JWST and a 4m UV/optical space telescope will provide our first observations of the oldest MSTO outside the LG to $\sim$2-3~Mpc.  However, truly transformative science requires the angular resolution, sensitivity, and stability of a $\sim$9-15m optical space telescope.  Observations of the oldest MSTO will also enable precise transverse motions for resolved stars in and around the LG and for bulk motions of galaxies out to several Mpc, given sufficient time baselines. 
\vspace{0.15cm}

\noindent \textbf{HST}:~~HST-based science remains at the forefront of astrophysics, including near-field cosmology. 
Without HST, US astronomy will lack a UV/optical observatory in space for the 15-20 years it will take to develop and launch a next-generation facility.  
We strongly advocate for serious exploration of a servicing mission (public or private) to extend the lifetime of HST beyond 2025.  Such a mission may prove to be a cost effective way to mitigate $\sim15$ years or more of no high-angular resolution UV/optical space-based capabilities between the planned decommissioning of HST and the launch of a potential replacement.
\vspace{0.15cm}

\noindent \textbf{Further Scientific Synergy}:  Though this white paper is broad, it is not an exhaustive exploration of near-field cosmology.  For example, precise spatially resolved HI rotation curves of dwarf galaxies would provide deep insight into the nature of dark matter. A space-based UV spectograph would revolutionize our knowledge of the circumgalactic medium and baryon cycle of low-mass galaxies\cite{tumlinson2017}, while study of the hot interstellar and circumgalactic medium via X-rays is critical to understanding stellar feedback\cite{lynx2018}.

\vspace{0.15cm}
\noindent High resolution and alternative versions of the figures can be found \href{https://github.com/dweisz/near-field_Astro2020}{\nolinkurl{here}}.

\newpage
\bibliography{main.bbl}

\begin{thebibliography}{10}
\expandafter\ifx\csname url\endcsname\relax
  \def\url#1{\texttt{#1}}\fi
\expandafter\ifx\csname urlprefix\endcsname\relax\def\urlprefix{URL }\fi
\providecommand{\bibinfo}[2]{#2}
\providecommand{\eprint}[2][]{\url{#2}}

\bibitem{bullock2017}
\bibinfo{author}{{Bullock}, J.~S.} \& \bibinfo{author}{{Boylan-Kolchin}, M.}
\newblock \bibinfo{title}{{Small-Scale Challenges to the {$\Lambda$}CDM
  Paradigm}}.
\newblock \emph{\bibinfo{journal}{\araa}} \textbf{\bibinfo{volume}{55}},
  \bibinfo{pages}{343--387} (\bibinfo{year}{2017}).

\bibitem{buckley2018}
\bibinfo{author}{{Buckley}, M.~R.} \& \bibinfo{author}{{Peter}, A.~H.~G.}
\newblock \bibinfo{title}{{Gravitational probes of dark matter physics}}.
\newblock \emph{\bibinfo{journal}{\physrep}} \textbf{\bibinfo{volume}{761}},
  \bibinfo{pages}{1--60} (\bibinfo{year}{2018}).

\bibitem{jethwa2018}
\bibinfo{author}{{Jethwa}, P.}, \bibinfo{author}{{Erkal}, D.} \&
  \bibinfo{author}{{Belokurov}, V.}
\newblock \bibinfo{title}{{The upper bound on the lowest mass halo}}.
\newblock \emph{\bibinfo{journal}{\mnras}} \textbf{\bibinfo{volume}{473}},
  \bibinfo{pages}{2060--2083} (\bibinfo{year}{2018}).

\bibitem{tollerud2008}
\bibinfo{author}{{Tollerud}, E.~J.}, \bibinfo{author}{{Bullock}, J.~S.},
  \bibinfo{author}{{Strigari}, L.~E.} \& \bibinfo{author}{{Willman}, B.}
\newblock \bibinfo{title}{{Hundreds of Milky Way Satellites? Luminosity Bias in
  the Satellite Luminosity Function}}.
\newblock \emph{\bibinfo{journal}{\apj}} \textbf{\bibinfo{volume}{688}},
  \bibinfo{pages}{277--289} (\bibinfo{year}{2008}).

\bibitem{kim2018}
\bibinfo{author}{{Kim}, S.~Y.}, \bibinfo{author}{{Peter}, A.~H.~G.} \&
  \bibinfo{author}{{Hargis}, J.~R.}
\newblock \bibinfo{title}{{Missing Satellites Problem: Completeness Corrections
  to the Number of Satellite Galaxies in the Milky Way are Consistent with Cold
  Dark Matter Predictions}}.
\newblock \emph{\bibinfo{journal}{Physical Review Letters}}
  \textbf{\bibinfo{volume}{121}}, \bibinfo{pages}{211302}
  (\bibinfo{year}{2018}).

\bibitem{kelley2018}
\bibinfo{author}{{Kelley}, T.} \emph{et~al.}
\newblock \bibinfo{title}{{Phat ELVIS: The inevitable effect of the Milky Way's
  disk on its dark matter subhaloes}}.
\newblock \emph{\bibinfo{journal}{{arXiv:1811.12413 [astro-ph]}}}
  (\bibinfo{year}{2018}).

\bibitem{brooks2014}
\bibinfo{author}{{Brooks}, A.~M.} \& \bibinfo{author}{{Zolotov}, A.}
\newblock \bibinfo{title}{{Why Baryons Matter: The Kinematics of Dwarf
  Spheroidal Satellites}}.
\newblock \emph{\bibinfo{journal}{\apj}} \textbf{\bibinfo{volume}{786}},
  \bibinfo{pages}{87} (\bibinfo{year}{2014}).

\bibitem{garrison-kimmel2017}
\bibinfo{author}{{Garrison-Kimmel}, S.} \emph{et~al.}
\newblock \bibinfo{title}{{Not so lumpy after all: modelling the depletion of
  dark matter subhaloes by Milky Way-like galaxies}}.
\newblock \emph{\bibinfo{journal}{\mnras}} \textbf{\bibinfo{volume}{471}},
  \bibinfo{pages}{1709--1727} (\bibinfo{year}{2017}).

\bibitem{carlberg2009a}
\bibinfo{author}{{Carlberg}, R.~G.}
\newblock \bibinfo{title}{{Star Stream Folding by Dark Galactic Subhalos}}.
\newblock \emph{\bibinfo{journal}{\apjl}} \textbf{\bibinfo{volume}{705}},
  \bibinfo{pages}{L223--L226} (\bibinfo{year}{2009}).

\bibitem{yoon2011}
\bibinfo{author}{{Yoon}, J.~H.}, \bibinfo{author}{{Johnston}, K.~V.} \&
  \bibinfo{author}{{Hogg}, D.~W.}
\newblock \bibinfo{title}{{Clumpy Streams from Clumpy Halos: Detecting Missing
  Satellites with Cold Stellar Structures}}.
\newblock \emph{\bibinfo{journal}{\apj}} \textbf{\bibinfo{volume}{731}},
  \bibinfo{pages}{58} (\bibinfo{year}{2011}).

\bibitem{penarrubia2016}
\bibinfo{author}{{Pe{\~n}arrubia}, J.}, \bibinfo{author}{{Ludlow}, A.~D.},
  \bibinfo{author}{{Chanam{\'e}}, J.} \& \bibinfo{author}{{Walker}, M.~G.}
\newblock \bibinfo{title}{{Wide binaries in ultrafaint galaxies: a window on to
  dark matter on the smallest scales}}.
\newblock \emph{\bibinfo{journal}{\mnras}} \textbf{\bibinfo{volume}{461}},
  \bibinfo{pages}{L72--L76} (\bibinfo{year}{2016}).

\bibitem{dalal2002}
\bibinfo{author}{{Dalal}, N.} \& \bibinfo{author}{{Kochanek}, C.~S.}
\newblock \bibinfo{title}{{Direct Detection of Cold Dark Matter Substructure}}.
\newblock \emph{\bibinfo{journal}{\apj}} \textbf{\bibinfo{volume}{572}},
  \bibinfo{pages}{25--33} (\bibinfo{year}{2002}).

\bibitem{vegetti2010}
\bibinfo{author}{{Vegetti}, S.}, \bibinfo{author}{{Koopmans}, L.~V.~E.},
  \bibinfo{author}{{Bolton}, A.}, \bibinfo{author}{{Treu}, T.} \&
  \bibinfo{author}{{Gavazzi}, R.}
\newblock \bibinfo{title}{{Detection of a dark substructure through
  gravitational imaging}}.
\newblock \emph{\bibinfo{journal}{\mnras}} \textbf{\bibinfo{volume}{408}},
  \bibinfo{pages}{1969--1981} (\bibinfo{year}{2010}).

\bibitem{dicintio2014}
\bibinfo{author}{{Di Cintio}, A.} \emph{et~al.}
\newblock \bibinfo{title}{{A mass-dependent density profile for dark matter
  haloes including the influence of galaxy formation}}.
\newblock \emph{\bibinfo{journal}{\mnras}} \textbf{\bibinfo{volume}{441}},
  \bibinfo{pages}{2986--2995} (\bibinfo{year}{2014}).

\bibitem{onorbe2015}
\bibinfo{author}{{O{\~n}orbe}, J.} \emph{et~al.}
\newblock \bibinfo{title}{{Forged in FIRE: cusps, cores and baryons in low-mass
  dwarf galaxies}}.
\newblock \emph{\bibinfo{journal}{\mnras}} \textbf{\bibinfo{volume}{454}},
  \bibinfo{pages}{2092--2106} (\bibinfo{year}{2015}).

\bibitem{read2016}
\bibinfo{author}{{Read}, J.~I.}, \bibinfo{author}{{Agertz}, O.} \&
  \bibinfo{author}{{Collins}, M.~L.~M.}
\newblock \bibinfo{title}{{Dark matter cores all the way down}}.
\newblock \emph{\bibinfo{journal}{\mnras}} \textbf{\bibinfo{volume}{459}},
  \bibinfo{pages}{2573--2590} (\bibinfo{year}{2016}).

\bibitem{dekel1986}
\bibinfo{author}{{Dekel}, A.} \& \bibinfo{author}{{Silk}, J.}
\newblock \bibinfo{title}{{The origin of dwarf galaxies, cold dark matter, and
  biased galaxy formation}}.
\newblock \emph{\bibinfo{journal}{\apj}} \textbf{\bibinfo{volume}{303}},
  \bibinfo{pages}{39--55} (\bibinfo{year}{1986}).

\bibitem{tollet2016}
\bibinfo{author}{{Tollet}, E.} \emph{et~al.}
\newblock \bibinfo{title}{{NIHAO - IV: core creation and destruction in dark
  matter density profiles across cosmic time}}.
\newblock \emph{\bibinfo{journal}{\mnras}} \textbf{\bibinfo{volume}{456}},
  \bibinfo{pages}{3542--3552} (\bibinfo{year}{2016}).

\bibitem{fitts2017}
\bibinfo{author}{{Fitts}, A.} \emph{et~al.}
\newblock \bibinfo{title}{{fire in the field: simulating the threshold of
  galaxy formation}}.
\newblock \emph{\bibinfo{journal}{\mnras}} \textbf{\bibinfo{volume}{471}},
  \bibinfo{pages}{3547--3562} (\bibinfo{year}{2017}).

\bibitem{read2019}
\bibinfo{author}{{Read}, J.~I.}, \bibinfo{author}{{Walker}, M.~G.} \&
  \bibinfo{author}{{Steger}, P.}
\newblock \bibinfo{title}{{Dark matter heats up in dwarf galaxies}}.
\newblock \emph{\bibinfo{journal}{{arXiv:1808.06634 [astro-ph]}}}
  (\bibinfo{year}{2019}).

\bibitem{simon2019}
\bibinfo{author}{{Simon}, J.~D.}
\newblock \bibinfo{title}{{The Faintest Dwarf Galaxies}}.
\newblock \emph{\bibinfo{journal}{{arXiv:1901.05465 [astro-ph]}}}
  (\bibinfo{year}{2019}).

\bibitem{tosi1989}
\bibinfo{author}{{Tosi}, M.}, \bibinfo{author}{{Greggio}, L.} \&
  \bibinfo{author}{{Focardi}, P.}
\newblock \bibinfo{title}{{Star formation in dwarf irregular galaxies -
  Preliminary results}}.
\newblock \emph{\bibinfo{journal}{\apss}} \textbf{\bibinfo{volume}{156}},
  \bibinfo{pages}{295--300} (\bibinfo{year}{1989}).

\bibitem{tolstoy1996}
\bibinfo{author}{{Tolstoy}, E.}
\newblock \bibinfo{title}{{The Resolved Stellar Population of Leo A}}.
\newblock \emph{\bibinfo{journal}{\apj}} \textbf{\bibinfo{volume}{462}},
  \bibinfo{pages}{684} (\bibinfo{year}{1996}).

\bibitem{dolphin2002}
\bibinfo{author}{{Dolphin}, A.~E.}
\newblock \bibinfo{title}{{Numerical methods of star formation history
  measurement and applications to seven dwarf spheroidals}}.
\newblock \emph{\bibinfo{journal}{MNRAS}} \textbf{\bibinfo{volume}{332}},
  \bibinfo{pages}{91--108} (\bibinfo{year}{2002}).

\bibitem{skillman2002}
\bibinfo{author}{{Skillman}, E.~D.} \& \bibinfo{author}{{Gallart}, C.}
\newblock \bibinfo{title}{{First Results of the Coimbra Experiment}}.
\newblock In \bibinfo{editor}{{Lejeune}, T.} \& \bibinfo{editor}{{Fernandes},
  J.} (eds.) \emph{\bibinfo{booktitle}{Observed HR Diagrams and Stellar
  Evolution}}, vol. \bibinfo{volume}{274} of
  \emph{\bibinfo{series}{Astronomical Society of the Pacific Conference
  Series}}, \bibinfo{pages}{535} (\bibinfo{year}{2002}).

\bibitem{brown2004b}
\bibinfo{author}{{Brown}, T.~M.} \emph{et~al.}
\newblock \bibinfo{title}{{Age Constraints for an M31 Globular Cluster from
  Main-Sequence Photometry}}.
\newblock \emph{\bibinfo{journal}{\apjl}} \textbf{\bibinfo{volume}{613}},
  \bibinfo{pages}{L125--L128} (\bibinfo{year}{2004}).

\bibitem{hidalgo2009}
\bibinfo{author}{{Hidalgo}, S.~L.}, \bibinfo{author}{{Aparicio}, A.},
  \bibinfo{author}{{Mart{\'{\i}}nez-Delgado}, D.} \&
  \bibinfo{author}{{Gallart}, C.}
\newblock \bibinfo{title}{{On the Extended Structure of the Phoenix Dwarf
  Galaxy}}.
\newblock \emph{\bibinfo{journal}{\apj}} \textbf{\bibinfo{volume}{705}},
  \bibinfo{pages}{704--716} (\bibinfo{year}{2009}).

\bibitem{monelli2010b}
\bibinfo{author}{{Monelli}, M.} \emph{et~al.}
\newblock \bibinfo{title}{{The ACS LCID Project. III. The Star Formation
  History of the Cetus dSph Galaxy: A Post-reionization Fossil}}.
\newblock \emph{\bibinfo{journal}{\apj}} \textbf{\bibinfo{volume}{720}},
  \bibinfo{pages}{1225--1245} (\bibinfo{year}{2010}).

\bibitem{cole2014}
\bibinfo{author}{{Cole}, A.~A.} \emph{et~al.}
\newblock \bibinfo{title}{{Delayed Star Formation in Isolated Dwarf galaxies:
  Hubble Space Telescope Star Formation History of the Aquarius Dwarf
  Irregular}}.
\newblock \emph{\bibinfo{journal}{\apj}} \textbf{\bibinfo{volume}{795}},
  \bibinfo{pages}{54} (\bibinfo{year}{2014}).

\bibitem{gallart2015}
\bibinfo{author}{{Gallart}, C.} \emph{et~al.}
\newblock \bibinfo{title}{{The ACS LCID Project: On the Origin of Dwarf Galaxy
  Types---A Manifestation of the Halo Assembly Bias?}}
\newblock \emph{\bibinfo{journal}{\apjl}} \textbf{\bibinfo{volume}{811}},
  \bibinfo{pages}{L18} (\bibinfo{year}{2015}).

\bibitem{weisz2011a}
\bibinfo{author}{{Weisz}, D.~R.} \emph{et~al.}
\newblock \bibinfo{title}{{The ACS Nearby Galaxy Survey Treasury. VIII. The
  Global Star Formation Histories of 60 Dwarf Galaxies in the Local Volume}}.
\newblock \emph{\bibinfo{journal}{\apj}} \textbf{\bibinfo{volume}{739}},
  \bibinfo{pages}{5} (\bibinfo{year}{2011}).

\bibitem{dolphin2012}
\bibinfo{author}{{Dolphin}, A.~E.}
\newblock \bibinfo{title}{{On the Estimation of Systematic Uncertainties of
  Star Formation Histories}}.
\newblock \emph{\bibinfo{journal}{\apj}} \textbf{\bibinfo{volume}{751}},
  \bibinfo{pages}{60} (\bibinfo{year}{2012}).

\bibitem{brown2014}
\bibinfo{author}{{Brown}, T.~M.} \emph{et~al.}
\newblock \bibinfo{title}{{The Quenching of the Ultra-faint Dwarf Galaxies in
  the Reionization Era}}.
\newblock \emph{\bibinfo{journal}{\apj}} \textbf{\bibinfo{volume}{796}},
  \bibinfo{pages}{91} (\bibinfo{year}{2014}).

\bibitem{weisz2014a}
\bibinfo{author}{{Weisz}, D.~R.} \emph{et~al.}
\newblock \bibinfo{title}{{The Star Formation Histories of Local Group Dwarf
  Galaxies. I. Hubble Space Telescope/Wide Field Planetary Camera 2
  Observations}}.
\newblock \emph{\bibinfo{journal}{\apj}} \textbf{\bibinfo{volume}{789}},
  \bibinfo{pages}{147} (\bibinfo{year}{2014}).

\bibitem{skillman2017}
\bibinfo{author}{{Skillman}, E.~D.} \emph{et~al.}
\newblock \bibinfo{title}{{The ISLAndS Project. II. The Lifetime Star Formation
  Histories of Six Andomeda dSphS}}.
\newblock \emph{\bibinfo{journal}{\apj}} \textbf{\bibinfo{volume}{837}},
  \bibinfo{pages}{102} (\bibinfo{year}{2017}).

\bibitem{skillman2014}
\bibinfo{author}{{Skillman}, E.~D.} \emph{et~al.}
\newblock \bibinfo{title}{{The ACS LCID Project. X. The Star Formation History
  of IC~1613: Revisiting the Over-cooling Problem}}.
\newblock \emph{\bibinfo{journal}{\apj}} \textbf{\bibinfo{volume}{786}},
  \bibinfo{pages}{44} (\bibinfo{year}{2014}).

\bibitem{sohn2013}
\bibinfo{author}{{Sohn}, S.~T.} \emph{et~al.}
\newblock \bibinfo{title}{{The Space Motion of Leo I: Hubble Space Telescope
  Proper Motion and Implied Orbit}}.
\newblock \emph{\bibinfo{journal}{\apj}} \textbf{\bibinfo{volume}{768}},
  \bibinfo{pages}{139} (\bibinfo{year}{2013}).

\bibitem{wetzel2015c}
\bibinfo{author}{{Wetzel}, A.~R.}, \bibinfo{author}{{Tollerud}, E.~J.} \&
  \bibinfo{author}{{Weisz}, D.~R.}
\newblock \bibinfo{title}{{Rapid Environmental Quenching of Satellite Dwarf
  Galaxies in the Local Group}}.
\newblock \emph{\bibinfo{journal}{\apjl}} \textbf{\bibinfo{volume}{808}},
  \bibinfo{pages}{L27} (\bibinfo{year}{2015}).

\bibitem{fillingham2016}
\bibinfo{author}{{Fillingham}, S.~P.} \emph{et~al.}
\newblock \bibinfo{title}{{Under pressure: quenching star formation in low-mass
  satellite galaxies via stripping}}.
\newblock \emph{\bibinfo{journal}{\mnras}} \textbf{\bibinfo{volume}{463}},
  \bibinfo{pages}{1916--1928} (\bibinfo{year}{2016}).

\bibitem{kallivayalil2015}
\bibinfo{author}{{Kallivayalil}, N.} \emph{et~al.}
\newblock \bibinfo{title}{{A Hubble Astrometry Initiative: Laying the
  Foundation for the Next-Generation Proper-Motion Survey of the Local Group}}.
\newblock \emph{\bibinfo{journal}{{arXiv:1503.01785 [astro-ph]}}}
  (\bibinfo{year}{2015}).

\bibitem{fritz2018}
\bibinfo{author}{{Fritz}, T.~K.} \emph{et~al.}
\newblock \bibinfo{title}{{Gaia DR2 proper motions of dwarf galaxies within 420
  kpc. Orbits, Milky Way mass, tidal influences, planar alignments, and group
  infall}}.
\newblock \emph{\bibinfo{journal}{\aap}} \textbf{\bibinfo{volume}{619}},
  \bibinfo{pages}{A103} (\bibinfo{year}{2018}).

\bibitem{simon2018}
\bibinfo{author}{{Simon}, J.~D.}
\newblock \bibinfo{title}{{Gaia Proper Motions and Orbits of the Ultra-faint
  Milky Way Satellites}}.
\newblock \emph{\bibinfo{journal}{\apj}} \textbf{\bibinfo{volume}{863}},
  \bibinfo{pages}{89} (\bibinfo{year}{2018}).

\bibitem{cole2007}
\bibinfo{author}{{Cole}, A.~A.} \emph{et~al.}
\newblock \bibinfo{title}{{Leo A: A Late-blooming Survivor of the Epoch of
  Reionization in the Local Group}}.
\newblock \emph{\bibinfo{journal}{\apjl}} \textbf{\bibinfo{volume}{659}},
  \bibinfo{pages}{L17--L20} (\bibinfo{year}{2007}).

\bibitem{mcconnachie2012}
\bibinfo{author}{{McConnachie}, A.~W.}
\newblock \bibinfo{title}{{The Observed Properties of Dwarf Galaxies in and
  around the Local Group}}.
\newblock \emph{\bibinfo{journal}{\aj}} \textbf{\bibinfo{volume}{144}},
  \bibinfo{pages}{4} (\bibinfo{year}{2012}).

\bibitem{dolphin2013}
\bibinfo{author}{{Dolphin}, A.~E.}
\newblock \bibinfo{title}{{On the Estimation of Random Uncertainties of Star
  Formation Histories}}.
\newblock \emph{\bibinfo{journal}{\apj}} \textbf{\bibinfo{volume}{775}},
  \bibinfo{pages}{76} (\bibinfo{year}{2013}).

\bibitem{Robertson:2010vq}
\bibinfo{author}{{Robertson}, B.~E.}, \bibinfo{author}{{Ellis}, R.~S.},
  \bibinfo{author}{{Dunlop}, J.~S.}, \bibinfo{author}{{McLure}, R.~J.} \&
  \bibinfo{author}{{Stark}, D.~P.}
\newblock \bibinfo{title}{{Early star-forming galaxies and the reionization of
  the Universe}}.
\newblock \emph{\bibinfo{journal}{\nat}} \textbf{\bibinfo{volume}{468}},
  \bibinfo{pages}{49--55} (\bibinfo{year}{2010}).

\bibitem{robertson2015}
\bibinfo{author}{{Robertson}, B.~E.}, \bibinfo{author}{{Ellis}, R.~S.},
  \bibinfo{author}{{Furlanetto}, S.~R.} \& \bibinfo{author}{{Dunlop}, J.~S.}
\newblock \bibinfo{title}{{Cosmic Reionization and Early Star-forming Galaxies:
  A Joint Analysis of New Constraints from Planck and the Hubble Space
  Telescope}}.
\newblock \emph{\bibinfo{journal}{\apjl}} \textbf{\bibinfo{volume}{802}},
  \bibinfo{pages}{L19} (\bibinfo{year}{2015}).

\bibitem{mirocha2019}
\bibinfo{author}{{Mirocha}, J.} \& \bibinfo{author}{{Furlanetto}, S.~R.}
\newblock \bibinfo{title}{{What does the first highly redshifted 21-cm
  detection tell us about early galaxies?}}
\newblock \emph{\bibinfo{journal}{\mnras}} \textbf{\bibinfo{volume}{483}},
  \bibinfo{pages}{1980--1992} (\bibinfo{year}{2019}).

\bibitem{weisz2014d}
\bibinfo{author}{{Weisz}, D.~R.}, \bibinfo{author}{{Johnson}, B.~D.} \&
  \bibinfo{author}{{Conroy}, C.}
\newblock \bibinfo{title}{{The Very Faint End of the UV Luminosity Function
  over Cosmic Time: Constraints from the Local Group Fossil Record}}.
\newblock \emph{\bibinfo{journal}{\apjl}} \textbf{\bibinfo{volume}{794}},
  \bibinfo{pages}{L3} (\bibinfo{year}{2014}).

\bibitem{boylankolchin2015}
\bibinfo{author}{{Boylan-Kolchin}, M.} \emph{et~al.}
\newblock \bibinfo{title}{{The Local Group as a time machine: studying the
  high-redshift Universe with nearby galaxies}}.
\newblock \emph{\bibinfo{journal}{\mnras}} \textbf{\bibinfo{volume}{453}},
  \bibinfo{pages}{1503--1512} (\bibinfo{year}{2015}).

\bibitem{weisz2017}
\bibinfo{author}{{Weisz}, D.~R.} \& \bibinfo{author}{{Boylan-Kolchin}, M.}
\newblock \bibinfo{title}{{Local Group ultra-faint dwarf galaxies in the
  reionization era}}.
\newblock \emph{\bibinfo{journal}{\mnras}} \textbf{\bibinfo{volume}{469}},
  \bibinfo{pages}{L83--L88} (\bibinfo{year}{2017}).

\bibitem{bullock2000}
\bibinfo{author}{{Bullock}, J.~S.}, \bibinfo{author}{{Kravtsov}, A.~V.} \&
  \bibinfo{author}{{Weinberg}, D.~H.}
\newblock \bibinfo{title}{{Reionization and the Abundance of Galactic
  Satellites}}.
\newblock \emph{\bibinfo{journal}{\apj}} \textbf{\bibinfo{volume}{539}},
  \bibinfo{pages}{517--521} (\bibinfo{year}{2000}).

\bibitem{benson2002}
\bibinfo{author}{{Benson}, A.~J.}, \bibinfo{author}{{Frenk}, C.~S.},
  \bibinfo{author}{{Lacey}, C.~G.}, \bibinfo{author}{{Baugh}, C.~M.} \&
  \bibinfo{author}{{Cole}, S.}
\newblock \bibinfo{title}{{The effects of photoionization on galaxy formation -
  II. Satellite galaxies in the Local Group}}.
\newblock \emph{\bibinfo{journal}{\mnras}} \textbf{\bibinfo{volume}{333}},
  \bibinfo{pages}{177--190} (\bibinfo{year}{2002}).

\bibitem{klypin1999}
\bibinfo{author}{{Klypin}, A.}, \bibinfo{author}{{Kravtsov}, A.~V.},
  \bibinfo{author}{{Valenzuela}, O.} \& \bibinfo{author}{{Prada}, F.}
\newblock \bibinfo{title}{{Where Are the Missing Galactic Satellites?}}
\newblock \emph{\bibinfo{journal}{\apj}} \textbf{\bibinfo{volume}{522}},
  \bibinfo{pages}{82--92} (\bibinfo{year}{1999}).

\bibitem{moore1999}
\bibinfo{author}{{Moore}, B.} \emph{et~al.}
\newblock \bibinfo{title}{{Dark Matter Substructure within Galactic Halos}}.
\newblock \emph{\bibinfo{journal}{\apjl}} \textbf{\bibinfo{volume}{524}},
  \bibinfo{pages}{L19--L22} (\bibinfo{year}{1999}).

\bibitem{ocvirk2013}
\bibinfo{author}{{Ocvirk}, P.} \emph{et~al.}
\newblock \bibinfo{title}{{High-resolution Simulations of the Reionization of
  an Isolated Milky Way-M31 Galaxy Pair}}.
\newblock \emph{\bibinfo{journal}{\apj}} \textbf{\bibinfo{volume}{777}},
  \bibinfo{pages}{51} (\bibinfo{year}{2013}).

\bibitem{weisz2014m31}
\bibinfo{author}{{Weisz}, D.~R.} \emph{et~al.}
\newblock \bibinfo{title}{{Comparing M31 and Milky Way Satellites: The Extended
  Star Formation Histories of Andromeda II and Andromeda XVI}}.
\newblock \emph{\bibinfo{journal}{\apj}} \textbf{\bibinfo{volume}{789}},
  \bibinfo{pages}{24} (\bibinfo{year}{2014}).

\bibitem{monelli2016}
\bibinfo{author}{{Monelli}, M.} \emph{et~al.}
\newblock \bibinfo{title}{{The ISLANDS Project. I. Andromeda XVI, An Extremely
  Low Mass Galaxy Not Quenched by Reionization}}.
\newblock \emph{\bibinfo{journal}{\apj}} \textbf{\bibinfo{volume}{819}},
  \bibinfo{pages}{147} (\bibinfo{year}{2016}).

\bibitem{ji2016}
\bibinfo{author}{{Ji}, A.~P.}, \bibinfo{author}{{Frebel}, A.},
  \bibinfo{author}{{Chiti}, A.} \& \bibinfo{author}{{Simon}, J.~D.}
\newblock \bibinfo{title}{{R-process enrichment from a single event in an
  ancient dwarf galaxy}}.
\newblock \emph{\bibinfo{journal}{\nat}} \textbf{\bibinfo{volume}{531}},
  \bibinfo{pages}{610--613} (\bibinfo{year}{2016}).

\bibitem{duggan2018}
\bibinfo{author}{{Duggan}, G.~E.}, \bibinfo{author}{{Kirby}, E.~N.},
  \bibinfo{author}{{Andrievsky}, S.~M.} \& \bibinfo{author}{{Korotin}, S.~A.}
\newblock \bibinfo{title}{{Neutron Star Mergers are the Dominant Source of the
  r-process in the Early Evolution of Dwarf Galaxies}}.
\newblock \emph{\bibinfo{journal}{\apj}} \textbf{\bibinfo{volume}{869}},
  \bibinfo{pages}{50} (\bibinfo{year}{2018}).

\bibitem{marshall2018}
\bibinfo{author}{{Marshall}, J.} \emph{et~al.}
\newblock \bibinfo{title}{{Chemical Abundance Analysis of Tucana III, the
  Second $r$-process Enhanced Ultra-Faint Dwarf Galaxy}}.
\newblock \emph{\bibinfo{journal}{{arXiv:1812.01022 [astro-ph]}}}
  (\bibinfo{year}{2018}).

\bibitem{kirby2009}
\bibinfo{author}{{Kirby}, E.~N.}, \bibinfo{author}{{Guhathakurta}, P.},
  \bibinfo{author}{{Bolte}, M.}, \bibinfo{author}{{Sneden}, C.} \&
  \bibinfo{author}{{Geha}, M.~C.}
\newblock \bibinfo{title}{{Multi-element Abundance Measurements from
  Medium-resolution Spectra. I. The Sculptor Dwarf Spheroidal Galaxy}}.
\newblock \emph{\bibinfo{journal}{\apj}} \textbf{\bibinfo{volume}{705}},
  \bibinfo{pages}{328--346} (\bibinfo{year}{2009}).

\bibitem{kirby2013}
\bibinfo{author}{{Kirby}, E.~N.} \emph{et~al.}
\newblock \bibinfo{title}{{The Universal Stellar Mass-Stellar Metallicity
  Relation for Dwarf Galaxies}}.
\newblock \emph{\bibinfo{journal}{\apj}} \textbf{\bibinfo{volume}{779}},
  \bibinfo{pages}{102} (\bibinfo{year}{2013}).

\bibitem{frebel2015}
\bibinfo{author}{{Frebel}, A.} \& \bibinfo{author}{{Norris}, J.~E.}
\newblock \bibinfo{title}{{Near-Field Cosmology with Extremely Metal-Poor
  Stars}}.
\newblock \emph{\bibinfo{journal}{\araa}} \textbf{\bibinfo{volume}{53}},
  \bibinfo{pages}{631--688} (\bibinfo{year}{2015}).

\bibitem{kirby2008a}
\bibinfo{author}{{Kirby}, E.~N.}, \bibinfo{author}{{Guhathakurta}, P.} \&
  \bibinfo{author}{{Sneden}, C.}
\newblock \bibinfo{title}{{Metallicity and Alpha-Element Abundance Measurement
  in Red Giant Stars from Medium-Resolution Spectra}}.
\newblock \emph{\bibinfo{journal}{\apj}} \textbf{\bibinfo{volume}{682}},
  \bibinfo{pages}{1217--1233} (\bibinfo{year}{2008}).

\bibitem{kirby2017}
\bibinfo{author}{{Kirby}, E.~N.} \emph{et~al.}
\newblock \bibinfo{title}{{Chemistry and Kinematics of the Late-forming Dwarf
  Irregular Galaxies Leo A, Aquarius, and Sagittarius DIG}}.
\newblock \emph{\bibinfo{journal}{\apj}} \textbf{\bibinfo{volume}{834}},
  \bibinfo{pages}{9} (\bibinfo{year}{2017}).

\bibitem{tolstoy2009}
\bibinfo{author}{{Tolstoy}, E.}, \bibinfo{author}{{Hill}, V.} \&
  \bibinfo{author}{{Tosi}, M.}
\newblock \bibinfo{title}{{Star-Formation Histories, Abundances, and Kinematics
  of Dwarf Galaxies in the Local Group}}.
\newblock \emph{\bibinfo{journal}{\araa}} \textbf{\bibinfo{volume}{47}},
  \bibinfo{pages}{371--425} (\bibinfo{year}{2009}).

\bibitem{roederer2012}
\bibinfo{author}{{Roederer}, I.~U.} \emph{et~al.}
\newblock \bibinfo{title}{{New Hubble Space Telescope Observations of Heavy
  Elements in Four Metal-Poor Stars}}.
\newblock \emph{\bibinfo{journal}{The Astrophysical Journal Supplement Series}}
  \textbf{\bibinfo{volume}{203}}, \bibinfo{pages}{27} (\bibinfo{year}{2012}).

\bibitem{ho2016}
\bibinfo{author}{{Ho}, A.~Y.~Q.} \emph{et~al.}
\newblock \bibinfo{title}{{Survey Cross-Calibration with The Cannon:
  Apogee-scale Stellar Labels from Lamost Spectra}}.
\newblock \emph{\bibinfo{journal}{{arXiv:1602.00303 [astro-ph]}}}
  (\bibinfo{year}{2016}).

\bibitem{ting2017}
\bibinfo{author}{{Ting}, Y.-S.}, \bibinfo{author}{{Conroy}, C.},
  \bibinfo{author}{{Rix}, H.-W.} \& \bibinfo{author}{{Cargile}, P.}
\newblock \bibinfo{title}{{Prospects for Measuring Abundances of $>$20 Elements
  with Low-resolution Stellar Spectra}}.
\newblock \emph{\bibinfo{journal}{\apj}} \textbf{\bibinfo{volume}{843}},
  \bibinfo{pages}{32} (\bibinfo{year}{2017}).

\bibitem{keller2007}
\bibinfo{author}{{Keller}, S.~C.} \emph{et~al.}
\newblock \bibinfo{title}{{The SkyMapper Telescope and The Southern Sky
  Survey}}.
\newblock \emph{\bibinfo{journal}{Publications of the Astronomical Society of
  Australia}} \textbf{\bibinfo{volume}{24}}, \bibinfo{pages}{1--12}
  (\bibinfo{year}{2007}).

\bibitem{starkenburg2017}
\bibinfo{author}{{Starkenburg}, E.} \emph{et~al.}
\newblock \bibinfo{title}{{The Pristine survey - I. Mining the Galaxy for the
  most metal-poor stars}}.
\newblock \emph{\bibinfo{journal}{\mnras}} \textbf{\bibinfo{volume}{471}},
  \bibinfo{pages}{2587--2604} (\bibinfo{year}{2017}).

\bibitem{cui2012}
\bibinfo{author}{{Cui}, X.-Q.} \emph{et~al.}
\newblock \bibinfo{title}{{The Large Sky Area Multi-Object Fiber Spectroscopic
  Telescope (LAMOST)}}.
\newblock \emph{\bibinfo{journal}{Research in Astronomy and Astrophysics}}
  \textbf{\bibinfo{volume}{12}}, \bibinfo{pages}{1197--1242}
  (\bibinfo{year}{2012}).

\bibitem{chiba2016}
\bibinfo{author}{{Chiba}, M.}, \bibinfo{author}{{Cohen}, J.} \&
  \bibinfo{author}{{Wyse}, R. F.~G.}
\newblock \bibinfo{title}{{Galactic Archaeology with the Subaru Prime Focus
  Spectrograph}}.
\newblock In \bibinfo{editor}{{Bragaglia}, A.}, \bibinfo{editor}{{Arnaboldi},
  M.}, \bibinfo{editor}{{Rejkuba}, M.} \& \bibinfo{editor}{{Romano}, D.} (eds.)
  \emph{\bibinfo{booktitle}{The General Assembly of Galaxy Halos: Structure,
  Origin and Evolution}}, vol. \bibinfo{volume}{317} of
  \emph{\bibinfo{series}{IAU Symposium}}, \bibinfo{pages}{280--281}
  (\bibinfo{year}{2016}).

\bibitem{leitherer1999}
\bibinfo{author}{{Leitherer}, C.} \emph{et~al.}
\newblock \bibinfo{title}{{Starburst99: Synthesis Models for Galaxies with
  Active Star Formation}}.
\newblock \emph{\bibinfo{journal}{The Astrophysical Journal Supplement Series}}
  \textbf{\bibinfo{volume}{123}}, \bibinfo{pages}{3--40}
  (\bibinfo{year}{1999}).

\bibitem{erb2010}
\bibinfo{author}{{Erb}, D.~K.} \emph{et~al.}
\newblock \bibinfo{title}{{Physical Conditions in a Young, Unreddened,
  Low-metallicity Galaxy at High Redshift}}.
\newblock \emph{\bibinfo{journal}{\apj}} \textbf{\bibinfo{volume}{719}},
  \bibinfo{pages}{1168--1190} (\bibinfo{year}{2010}).

\bibitem{madau2014b}
\bibinfo{author}{{Madau}, P.} \& \bibinfo{author}{{Dickinson}, M.}
\newblock \bibinfo{title}{{Cosmic Star-Formation History}}.
\newblock \emph{\bibinfo{journal}{\araa}} \textbf{\bibinfo{volume}{52}},
  \bibinfo{pages}{415--486} (\bibinfo{year}{2014}).

\bibitem{stark2016}
\bibinfo{author}{{Stark}, D.~P.}
\newblock \bibinfo{title}{{Galaxies in the First Billion Years After the Big
  Bang}}.
\newblock \emph{\bibinfo{journal}{\araa}} \textbf{\bibinfo{volume}{54}},
  \bibinfo{pages}{761--803} (\bibinfo{year}{2016}).

\bibitem{senchyna2017}
\bibinfo{author}{{Senchyna}, P.} \emph{et~al.}
\newblock \bibinfo{title}{{Ultraviolet spectra of extreme nearby star-forming
  regions - approaching a local reference sample for JWST}}.
\newblock \emph{\bibinfo{journal}{\mnras}} \textbf{\bibinfo{volume}{472}},
  \bibinfo{pages}{2608--2632} (\bibinfo{year}{2017}).

\bibitem{berg2018}
\bibinfo{author}{{Berg}, D.~A.}, \bibinfo{author}{{Erb}, D.~K.},
  \bibinfo{author}{{Auger}, M.~W.}, \bibinfo{author}{{Pettini}, M.} \&
  \bibinfo{author}{{Brammer}, G.~B.}
\newblock \bibinfo{title}{{A Window on the Earliest Star Formation: Extreme
  Photoionization Conditions of a High-ionization, Low-metallicity Lensed
  Galaxy at $z\sim2$}}.
\newblock \emph{\bibinfo{journal}{\apj}} \textbf{\bibinfo{volume}{859}},
  \bibinfo{pages}{164} (\bibinfo{year}{2018}).

\bibitem{levesque2012}
\bibinfo{author}{{Levesque}, E.~M.}, \bibinfo{author}{{Leitherer}, C.},
  \bibinfo{author}{{Ekstrom}, S.}, \bibinfo{author}{{Meynet}, G.} \&
  \bibinfo{author}{{Schaerer}, D.}
\newblock \bibinfo{title}{{The Effects of Stellar Rotation. I. Impact on the
  Ionizing Spectra and Integrated Properties of Stellar Populations}}.
\newblock \emph{\bibinfo{journal}{\apj}} \textbf{\bibinfo{volume}{751}},
  \bibinfo{pages}{67} (\bibinfo{year}{2012}).

\bibitem{choi2017}
\bibinfo{author}{{Choi}, J.}, \bibinfo{author}{{Conroy}, C.} \&
  \bibinfo{author}{{Byler}, N.}
\newblock \bibinfo{title}{{The Evolution and Properties of Rotating Massive
  Star Populations}}.
\newblock \emph{\bibinfo{journal}{\apj}} \textbf{\bibinfo{volume}{838}},
  \bibinfo{pages}{159} (\bibinfo{year}{2017}).

\bibitem{stanway2016}
\bibinfo{author}{{Stanway}, E.~R.}, \bibinfo{author}{{Eldridge}, J.~J.} \&
  \bibinfo{author}{{Becker}, G.~D.}
\newblock \bibinfo{title}{{Stellar population effects on the inferred photon
  density at reionization}}.
\newblock \emph{\bibinfo{journal}{\mnras}} \textbf{\bibinfo{volume}{456}},
  \bibinfo{pages}{485--499} (\bibinfo{year}{2016}).

\bibitem{conroy2013}
\bibinfo{author}{{Conroy}, C.}
\newblock \bibinfo{title}{{Modeling the Panchromatic Spectral Energy
  Distributions of Galaxies}}.
\newblock \emph{\bibinfo{journal}{\araa}} \textbf{\bibinfo{volume}{51}},
  \bibinfo{pages}{393--455} (\bibinfo{year}{2013}).

\bibitem{steidel2016}
\bibinfo{author}{{Steidel}, C.~C.} \emph{et~al.}
\newblock \bibinfo{title}{{Reconciling the Stellar and Nebular Spectra of
  High-redshift Galaxies}}.
\newblock \emph{\bibinfo{journal}{\apj}} \textbf{\bibinfo{volume}{826}},
  \bibinfo{pages}{159} (\bibinfo{year}{2016}).

\bibitem{byler2017}
\bibinfo{author}{{Byler}, N.}, \bibinfo{author}{{Dalcanton}, J.~J.},
  \bibinfo{author}{{Conroy}, C.} \& \bibinfo{author}{{Johnson}, B.~D.}
\newblock \bibinfo{title}{{Nebular Continuum and Line Emission in Stellar
  Population Synthesis Models}}.
\newblock \emph{\bibinfo{journal}{\apj}} \textbf{\bibinfo{volume}{840}},
  \bibinfo{pages}{44} (\bibinfo{year}{2017}).

\bibitem{byler2018}
\bibinfo{author}{{Byler}, N.} \emph{et~al.}
\newblock \bibinfo{title}{{Stellar and Nebular Diagnostics in the Ultraviolet
  for Star-forming Galaxies}}.
\newblock \emph{\bibinfo{journal}{\apj}} \textbf{\bibinfo{volume}{863}},
  \bibinfo{pages}{14} (\bibinfo{year}{2018}).

\bibitem{garcia2014}
\bibinfo{author}{{Garcia}, M.}, \bibinfo{author}{{Herrero}, A.},
  \bibinfo{author}{{Najarro}, F.}, \bibinfo{author}{{Lennon}, D.~J.} \&
  \bibinfo{author}{{Alejandro Urbaneja}, M.}
\newblock \bibinfo{title}{{Winds of Low-metallicity OB-type Stars: HST-COS
  Spectroscopy in IC 1613}}.
\newblock \emph{\bibinfo{journal}{\apj}} \textbf{\bibinfo{volume}{788}},
  \bibinfo{pages}{64} (\bibinfo{year}{2014}).

\bibitem{garcia2019}
\bibinfo{author}{{Garcia}, M.}, \bibinfo{author}{{Herrero}, A.},
  \bibinfo{author}{{Najarro}, F.}, \bibinfo{author}{{Camacho}, I.} \&
  \bibinfo{author}{{Lorenzo}, M.}
\newblock \bibinfo{title}{{Ongoing star formation at the outskirts of Sextans
  A: spectroscopic detection of early O-type stars}}.
\newblock \emph{\bibinfo{journal}{\mnras}} \textbf{\bibinfo{volume}{484}},
  \bibinfo{pages}{422--430} (\bibinfo{year}{2019}).

\bibitem{evans2019}
\bibinfo{author}{{Evans}, C.~J.} \emph{et~al.}
\newblock \bibinfo{title}{{First stellar spectroscopy in Leo P}}.
\newblock \emph{\bibinfo{journal}{{arXiv:1901.01295 [astro-ph]}}}
  (\bibinfo{year}{2019}).

\bibitem{berg2012}
\bibinfo{author}{{Berg}, D.~A.} \emph{et~al.}
\newblock \bibinfo{title}{{Direct Oxygen Abundances for Low-luminosity LVL
  Galaxies}}.
\newblock \emph{\bibinfo{journal}{\apj}} \textbf{\bibinfo{volume}{754}},
  \bibinfo{pages}{98} (\bibinfo{year}{2012}).

\bibitem{heckman2011}
\bibinfo{author}{{Heckman}, T.~M.} \emph{et~al.}
\newblock \bibinfo{title}{{Extreme Feedback and the Epoch of Reionization:
  Clues in the Local Universe}}.
\newblock \emph{\bibinfo{journal}{\apj}} \textbf{\bibinfo{volume}{730}},
  \bibinfo{pages}{5} (\bibinfo{year}{2011}).

\bibitem{karachentsev2013}
\bibinfo{author}{{Karachentsev}, I.~D.}, \bibinfo{author}{{Makarov}, D.~I.} \&
  \bibinfo{author}{{Kaisina}, E.~I.}
\newblock \bibinfo{title}{{Updated Nearby Galaxy Catalog}}.
\newblock \emph{\bibinfo{journal}{\aj}} \textbf{\bibinfo{volume}{145}},
  \bibinfo{pages}{101} (\bibinfo{year}{2013}).

\bibitem{luvoir2018}
\bibinfo{author}{{The LUVOIR Team}}.
\newblock \bibinfo{title}{{The LUVOIR Mission Concept Study Interim Report}}.
\newblock \emph{\bibinfo{journal}{{arXiv:1809.09668 [astro-ph]}}}
  (\bibinfo{year}{2018}).

\bibitem{torrealba2016}
\bibinfo{author}{{Torrealba}, G.}, \bibinfo{author}{{Koposov}, S.~E.},
  \bibinfo{author}{{Belokurov}, V.} \& \bibinfo{author}{{Irwin}, M.}
\newblock \bibinfo{title}{{The feeble giant. Discovery of a large and diffuse
  Milky Way dwarf galaxy in the constellation of Crater}}.
\newblock \emph{\bibinfo{journal}{\mnras}} \textbf{\bibinfo{volume}{459}},
  \bibinfo{pages}{2370--2378} (\bibinfo{year}{2016}).

\bibitem{torrealba2018}
\bibinfo{author}{{Torrealba}, G.} \emph{et~al.}
\newblock \bibinfo{title}{{The hidden giant: discovery of an enormous Galactic
  dwarf satellite in Gaia DR2}}.
\newblock \emph{\bibinfo{journal}{arXiv e-prints}}
  \bibinfo{pages}{arXiv:1811.04082} (\bibinfo{year}{2018}).

\bibitem{mcconnachie2016}
\bibinfo{author}{{McConnachie}, A.} \emph{et~al.}
\newblock \bibinfo{title}{{The Detailed Science Case for the Maunakea
  Spectroscopic Explorer: the Composition and Dynamics of the Faint Universe}}.
\newblock \emph{\bibinfo{journal}{{arXiv:1606.00043 [astro-ph]}}}
  (\bibinfo{year}{2016}).

\bibitem{tumlinson2017}
\bibinfo{author}{{Tumlinson}, J.}, \bibinfo{author}{{Peeples}, M.~S.} \&
  \bibinfo{author}{{Werk}, J.~K.}
\newblock \bibinfo{title}{{The Circumgalactic Medium}}.
\newblock \emph{\bibinfo{journal}{Annual Review of Astronomy and Astrophysics}}
  \textbf{\bibinfo{volume}{55}}, \bibinfo{pages}{389--432}
  (\bibinfo{year}{2017}).

\bibitem{lynx2018}
\bibinfo{author}{{The Lynx Team}}.
\newblock \bibinfo{title}{{The Lynx Mission Concept Study Interim Report}}.
\newblock \emph{\bibinfo{journal}{arXiv e-prints}}
  \bibinfo{pages}{arXiv:1809.09642} (\bibinfo{year}{2018}).

\end{thebibliography}

\end{document}